\documentclass[referee]{raa}            % referee version: for submission

\usepackage{graphicx,times}             %for PS/EPS graphics inclusion, new
\usepackage{natbib}
\usepackage{amssymb,amsmath}

\usepackage{upgreek,multirow,makecell,diagbox,booktabs,color}%作者添加，公式和制表需要
\newcommand{\luoma}[1]{\uppercase\expandafter{\romannumeral#1}}%作者添加，将阿拉伯数字输出为罗马数字

\bibpunct{(}{)}{;}{a}{}{,}

\usepackage[a4paper=true,pagebackref=true]{hyperref}
\hypersetup{colorlinks = true, linkcolor = green, anchorcolor = red, citecolor = blue, filecolor = red, pagecolor = red, urlcolor = red}

\begin{document}
	
	\graphicspath{{figures/}}%所有eps图片在figures中
	
	\title{The estimate of sensitivity for large infrared telescopes based on measured sky brightness and atmospheric extinction
		%\,$^*$
		%\footnotetext{$*$ Supported by the National Natural Science Foundation of China.}
	}
	%   \subtitle{I. Place Your Subtitle Here}
	
	\volnopage{Vol.0 (20xx) No.0, 000--000}      %%preserved for Editor. DOn't remove!
	\setcounter{page}{1}          %%starting page, preserved for Editor. DOn't remove!
	
	\author{Zhi-Jun Zhao
		\inst{1,4}
		\and Hai-Jing Zhou
		\inst{1}
		\and Yu-Chen Zhang
		\inst{2}
		\and Yun Ling
		\inst{3} 
		\and Fang-Yu Xu$^*$
		\inst{2}
	}
	%% Here is an example of three authors come from different institutes.
	%% For single author or all the authors from an institute, use "\inst{}" only
	
	\institute{School of Physics, Henan Normal University,Xinxiang 453007, China; \\{\it xu\_fangyu@ynao.ac.cn; zhaozhijun@htu.edu.cn}\\
		%% Please give the E-mail address of the author, to whom future correspondence and
		%% offprint requests will be sent.
		\and
		Yunnan Observatories, Chinese Academy of Sciences, Kunming 650216, China;\\
		\and
		Kunming Institute of Physics, Kunming 650216, China;\\
		\and
		Henan Key Laboratory of Infrared Materials \& Spectrum Measures and Applications, Xinxiang 453007, China\\
		\vs\no
		{\small Received~~20xx month day; accepted~~20xx~~month day}}

\abstract{: In order to evaluate the ground-based infrared telescope sensitivity affected by the noise from the atmosphere, instruments and detectors, we construct a sensitivity model that can calculate limiting magnitudes and signal-to-noise ratio ($S/N$). The model is tested  with tentative measurements of $\rm M'$-band sky brightness and atmospheric extinction obtained at the Ali and Daocheng sites. We find that the noise caused by an excellent scientific detector and instruments at $-135^\circ \rm C$ can be ignored compared to the $\rm M'$-band sky background noise. Thus, when $S/N=3$  and total exposure time is 1 second for 10 m telescopes, the  magnitude limited by the atmosphere is $13.01^{\rm m}$ at Ali and  $12.96^{\rm m}$  at Daocheng. Even under less-than-ideal circumstances, i.e., the readout noise of a deep cryogenic detector is less than $200e^-$ and  the  instruments are cooled to below $-87.2^\circ \rm C$, the above magnitudes decrease by $0.056^{\rm m}$ at most. Therefore, according to observational requirements with a large telescope in a given infrared band, astronomers can use this sensitivity model as a tool for guiding site surveys, detector selection and instrumental thermal-control.
}

   \authorrunning{Zhi-Jun Zhao et al.  }            %author_head in even pages
   \titlerunning{The sensitivity of large infrared telescopes}  % title_head in odd pages

   \maketitle
%% The author head (on even pages) and the title head (on odd pages) will be
%% automatically extracted from \author{} and \title{}. Whenever the title is too long,
%% you will be asked to supply a shorter one by inserting either \authorrunning{} or
%% \titlerunning{} before \maketitle. Anyway, you can specify your own heads.
%%
%%
%% Note: In the following text body of your manuscript, please note several differences from
%%       other major journals:
%% (1) \subsection{Please Capitalize the First Letter of Each Notional Word in Subsection Title}
%% (2) Please Capitalize the First Letter of Each Notional Word in all tables' captions

%
%________________________________________________ sections below
%
\section{Introduction}           %% first-level sections will be auto-capitalized
\label{sect:intro}

Infrared observations have unique advantages in astronomical research such as observing cool
celestial objects and extra-solar planets. Many 8–10 m ground-based telescopes  (Keck, Gemini, Subaru, etc) are equipped with infrared instruments. Currently, a 12 m optical infrared telescope and an 8 m solar telescope scheduled to be built in China will be equipped with infrared instruments (\citealt{Cui2018,Deng2016,Liu2012}). On the ground, the sensitivity of large telescopes is affected by sky background, atmospheric extinction, detector noise, and instrumental  noise.

The sky background and atmospheric extinction set a fundamental sensitivity limit of infrared observations, so they are key indicators in site surveys of large infrared telescopes. From the 1970s to 1990s, the $1\upmu \rm m\sim 30\upmu \rm m $ infrared sky brightness was widely measured abroad
(\citealt{Westphal1974,Ashley1996,Smith1998,Phillips1999}). In China, site surveys have been carried out for many years on the dry and cold western plateau, and three excellent sites (Ali site in Tibet, Daocheng site in Sichuan, and Muztagh Ata in Xinjiang) have been listed as the candidates for large optical/infrared telescopes (\citealt{Yao2005,Wu2016,Song2020,Feng2020Site}). By continuously measuring air-temperature and water vapor content (\citealt{Qian2015Numberical,Wang2013,Liu2018}), astronomers find that the three sites may be suitable for infrared observations. Nevertheless, direct evidence that reflects the quality of sites used for infrared observations is scarce. The near-infrared (J, H and K band) sky brightness  at the Ali site has been measured with an InGaAs detector since 2017 (\citealt{Dong2018Design,Tang2018,Wang2018}). 
The $\rm M'$-band sky brightness and atmospheric transmittance were tentatively measured at the Daocheng site in March 2017 and Ali site in October 2017 (\citealt{Zhijun2017-phd,Wang2020}). In the thermal infrared such as $\rm M'$-band, even if the best sites for large telescopes have been indicated by measuring sky brightness and atmospheric extinction, it is also essential to strictly control the instrumental thermal-emission and detector noise in order to approach the fundamental sensitivity limit of infrared observations at the excellent sites. Hence, we need to study the various noise sources causing observational errors and determine the condition that the noise of the detectors and instruments can fulfill the required sensitivity of observations.

In this paper, we analyze the impact of various noise sources on the ground-based infrared telescope sensitivity, and then try to provide reasonable suggestions for detector selection and instrumental thermal-control. In Section (\ref{sect:obs_sensitivity}), we construct a sensitivity model that can be used to calculate the signal-to-noise ratio ($S/N$) and limiting magnitudes. In order to quantitatively describe the impact of noise caused by detectors and instruments on the $S/N$, we define a physical quantity called the quality factor ($Q_{snr}$) of the signal-to-noise ratio. In Section (\ref{sect:sky_ext}), we provide tentative measurements of $\rm M'$-band sky brightness and atmospheric extinction obtained at the Ali and Daocheng sites. In Section (\ref{sect:lm}) by using the above measurements, we calculate the $\rm M'$-band limiting magnitudes of 10 m telescopes and discuss the noise distribution and instrumental thermal-control. In Section (\ref{sect:conclusion}), final conclusions are summarized.

\section{The sensitivity model based on sky brightness and atmospheric extinction}
\label{sect:obs_sensitivity}

\subsection{The signal-to-noise ratio ($S/N$) and the exo-atmospheric magnitude}
\label{subsect:s2n}
In astronomical observations, noise is usually caused by fluctuations of electron numbers produced by a signal and background. The signal may be from a star, and the background generally consists of sky background, instrumental background, dark current, and readout noise. In order to find an accurate signal the background needs to be subtracted. However, in the thermal infrared the background is hard to accurately subtract because it usually fluctuates rapidly. Thus, the variance of the background noise may be added twice into the total noise variance (\citealt{Lena2012}).  Hence, the standard deviation of a fluctuating signal and background, which typically obeys a Poisson distribution, can be expressed as Equation (\ref{eq:sig_noise}):

\begin{equation}
\sigma=\sqrt{Sig\cdot t+2[B_{sky}\cdot t+B_{ins}\cdot t+n_{pixel}\cdot(Dark\cdot t +Ron^2)]},
\label{eq:sig_noise}
\end{equation}
where $Sig$ is the signal electrons per second, $B_{sky}$ is the electrons per second from the sky background, $B_{ins}$ is the electrons per second from the total instrumental background containing the emission from a telescope and relay optics, $t$ is the  elementary exposure time in seconds for the detector,  $Dark$ is the dark electrons per second at a unit pixel, $Ron$ is the RMS readout noise per pixel, and $n_{pixel}$ is the number of pixels covered by a star image.

The $S/N$ is shown in Equation (\ref{eq:s2n}):
\begin{align}
\frac{S}{N}=\frac{n_f\cdot Sig\cdot t}{\sqrt{n_f}\cdot \sigma}&=\sqrt{Sig\cdot n_f\cdot t}\sqrt{\frac{Sig\cdot t}{Sig\cdot t+2[B_{sky}\cdot t+B_{ins}\cdot t+n_{pixel}\cdot(Dark\cdot t+Ron^2)]}}\nonumber \\
&=\sqrt{n_f\cdot t}\sqrt{\frac{B_{sky}}{\beta}}\sqrt{\frac{1}{1+2(\beta+\beta\cdot \alpha_1+\beta\cdot\alpha_2)}},
\label{eq:s2n}
\end{align}
where $\beta=B_{sky}/Sig$, $\alpha_1=B_{ins}/B_{sky}$,  $\alpha_2=n_{pixel}\cdot(Dark\cdot t+Ron^2)/(B_{sky}\cdot t)$, $n_f$ is the number of superposed frames in multi-frame techniques, and $n_f\cdot t$ can be viewed as the total exposure time.

In Equation (\ref{eq:s2n}), the $Sig$, $B_{sky}$ and $B_{ins}$ in a given wave-band can be calculated  by

\begin{align}
Sig &=\tau_{atm}\tau_{ins}\eta F_{sig}  \pi(\frac{D}{2})^2 \frac{\lambda}{h c}
\label{eq:bgd_sig2},\\
B_{sky} &=\tau_{ins}\eta L_{sky} \pi(\frac{D}{2})^2 \Omega_{star}  \frac{\lambda}{h c}\label{eq:bgd_sig1},\\
B_{ins} &=\eta L_{ins} \pi(\frac{D}{2})^2 \Omega_{star}  \frac{\lambda}{h c}\label{eq:bgd_sig3}.
\end{align}
In Equations (\ref{eq:bgd_sig2},\ref{eq:bgd_sig1} and \ref{eq:bgd_sig3}), $\tau_{atm}$ is the atmospheric transmittance, $\tau_{ins}$ is the total instrumental transmittance, $\eta$ is the quantum efficiency of a detector, $F_{sig}$ is the flux of a star, $L_{sky}$ is the sky brightness, $L_{ins}$ is the radiance of instruments, $\Omega_{star}$ is the solid angle subtended by the image of a star, $D$ is the telescope diameter, $h$ is the Plank constant, $c$ is the velocity of light, and $\lambda$ is wavelength.

In Equations (\ref{eq:bgd_sig1} and \ref{eq:bgd_sig3}), the $L_{sky}$ and the $L_{ins}$ can be obtained by measurement or simulation. Based on the black-body radiation theory, for any object another form of radiance can also be expressed by 
\begin{equation}
L_{obj}=\epsilon_{obj}\cdot L_{obj}^b=\epsilon_{obj}^*\cdot L_{amb}^b,
\label{eq:eff_real}
\end{equation}
where $\epsilon_{obj}$ is the real emissivity, $\epsilon_{obj}^*$ is the effective emissivity, $L_{obj}$ is the radiance of an object, $L_{obj}^b$ is the black-body radiance at the object's temperature, and $L_{amb}^b$ is black-body radiance at the ambient temperature. By using the effective emissivity a more convenient form of $\alpha_1$ can be rewritten by

\begin{equation}
\alpha_1=\frac{B_{ins}}{B_{sky}}=\frac{\epsilon_{ins}^*\cdot L_{amb}^b}{\tau_{ins}\cdot \epsilon_{sky}^*\cdot L_{amb}^b}=\frac{\epsilon_{ins}^*}{\tau_{ins}\cdot \epsilon_{sky}^*},
\label{eq:alpha1}
\end{equation}
where $\epsilon_{sky}^*$ is  the effective emissivity of sky, and  $\epsilon_{ins}^*$ is the effective emissivity of the instruments.

From Equations (\ref{eq:bgd_sig2} and \ref{eq:bgd_sig1}), $F_{sig}$ can be expressed by

\begin{equation}
F_{sig}=\frac{L_{sky}\cdot\Omega_{star}}{\beta\cdot\tau_{atm}}.\label{eq:flux_3sigma}
\end{equation}
Then the exo-atmospheric magnitude of the stellar flux is obtained by
\begin{equation}
m_{sig}=-2.5{\rm log_{10}}(\frac{F_{sig}}{F_0\times \Delta\lambda})=-2.5{\rm log_{10}}(\frac{L_{sky}\cdot\Omega_{star}}{\beta\cdot\tau_{atm}}\cdot\frac{1}{F_0\times \Delta\lambda}),\label{eq:mag_3sigma}
\end{equation}
where $F_0$ is the spectral flux of zero magnitude in a given wave-band, and  $\Delta \lambda$ is the bandwidth of filter. 

Obviously, if the sky brightness and the atmospheric extinction at astronomical sites are given,  the relationship between $S/N$ or  $m_{sig}$ and $\beta$ can be easily obtained by Equation (\ref{eq:s2n} or \ref{eq:mag_3sigma}). Thus, for any $S/N$, $m_{sig}$ can be calculated.

\subsection{The quality factor of the signal-to-noise ratio}
\label{subsec:qsnr}
In the infrared observations on the ground, the detector noise and the instrument noise make the $S/N$ always lower than its theoretical limit (TL) determined by sky background and stellar intensity. In order to quantitatively describe the impact of noise caused by detectors and instruments on the $S/N$, it is necessary to introduce a physical quantity which can express the decrease of $S/N$. From the foregoing discussion  we find that when $(\alpha_1+\alpha_2)\rightarrow 0$,  $S/N$ reaches its theoretical limit. Thus, we obtain Equation (\ref{eq:s2ntl}): 
\begin{equation}
\frac{S}{N}\Big|_{TL}=\lim_{(\alpha_1+\alpha_2) \to 0}\frac{S}{N}=\sqrt{n_f\cdot t}\sqrt{\frac{B_{sky}}{\beta}}\sqrt{\frac{1}{1+2\beta}}.
\label{eq:s2ntl}
\end{equation}
According to Equations (\ref{eq:s2n} and \ref{eq:s2ntl}) a  physical quantity $Q_{snr}$ called the quality factor of S/N  can be defined as follows using Equation (\ref{eq:s2nratio}):

\begin{equation}
Q_{snr}=\frac{\frac{S}{N}}{\frac{S}{N}|_{TL}}=\frac{\sqrt{1+2\beta}}{\sqrt{1+2\beta+2\beta(\alpha_1+\alpha_2)}}.
\label{eq:s2nratio}
\end{equation}
Obviously, $0<Q_{snr}< 1$. The larger $Q_{snr}$ is, the better is the sensitivity of the telescope. 
For the  given $\beta$ and $Q_{snr}$ in Equation (\ref{eq:s2nratio}), $\alpha_1+\alpha_2$ can be calculated with Equation (\ref{eq:alphavalue}):
\begin{equation}
\alpha_1+\alpha_2 = (\frac{1}{Q_{snr}^2}-1)(1+\frac{1}{2\beta}).
\label{eq:alphavalue}
\end{equation}

From Equations (\ref{eq:s2ntl}, \ref{eq:s2nratio} and  \ref{eq:alphavalue}), one can obtain the following facts for a given sky background:

(1) For $2\beta\ll 1$, $S/N|_{TL}\approx\sqrt{n_f\cdot t\cdot B_{sky}/\beta}\propto \beta^{-1/2}$. The intensity of the stellar flux is far greater than the sky background, and the sensitivity is limited by signal shot noise. 

(2) For $2\beta\gg 1$, which is called the background-limited domain, $\alpha_1+\alpha_2\approx(Q_{snr}^{-2}-1)$ and $S/N|_{TL}\approx\sqrt{n_f\cdot t\cdot B_{sky}/2}/\beta\propto \beta^{-1}$. We take the partial derivative of $S/N|_{TL}$ with respect to $\beta$, and then obtain $\partial (S/N|_{TL})/(S/N|_{TL})=\partial\beta/\beta$. Likewise, taking the partial derivative of the $m_{sig}$ in Equation (\ref{eq:mag_3sigma}) with respect to $\beta$, we can obtain  $\partial  (m_{sig})={2.5}\times{\rm log_{10}}(e)\times\partial\beta/\beta$. Thus, in this domain, the slight reduction relative to the magnitude limited by sky background can be approximately obtained by
\begin{equation}
\Delta m_{sig} \approx{2.5}\times{\rm log_{10}}(e)\times\frac{\Delta(\frac{S}{N}|_{TL})}{\frac{S}{N}|_{TL}}=1.086\times(1-Q_{snr}).
\label{eq:mag_qsnr}
\end{equation}
Based on the desired $\Delta m_{sig}$, Equation (\ref{eq:mag_qsnr}) can be used to roughly estimate $Q_{snr}$. However, this only applies to the circumstances where $S/N$ approaches theoretical limit. By combining Equations (\ref{eq:alphavalue} and \ref{eq:mag_qsnr}), we can use $Q_{snr}$ to obtain reasonable requirements for detector noise and instrument noise. Thus, according to the $Q_{snr}$, the background-limited condition of astronomical sites (BLCAS) can be defined mathematically as
\begin{equation}
\left\{
\begin{split}
&2\beta\gg1 \\
&Q_{snr}\rightarrow 1
\end{split}
\right..
\label{eq:blcas}
\end{equation}
As illustrated in Equation (\ref{eq:mag_qsnr}), if we can accept that the  maximum magnitude reduction with respect to the magnitude limited by sky background is about $0.05^{\rm m}$, the minimum of $Q_{snr}$ should be about 0.95. Hence, in the background limited domain, the upper limit of $\alpha_1+\alpha_2$ is 0.11, which determines the maximum noise allowed from instruments and detectors.

\section{The measurements of $\rm M'$-band sky brightness and atmospheric extinction}
\label{sect:sky_ext}

\begin{figure}[htb]
	\begin{minipage}[t]{0.495\linewidth}
		\centering
		\includegraphics[width=70mm]{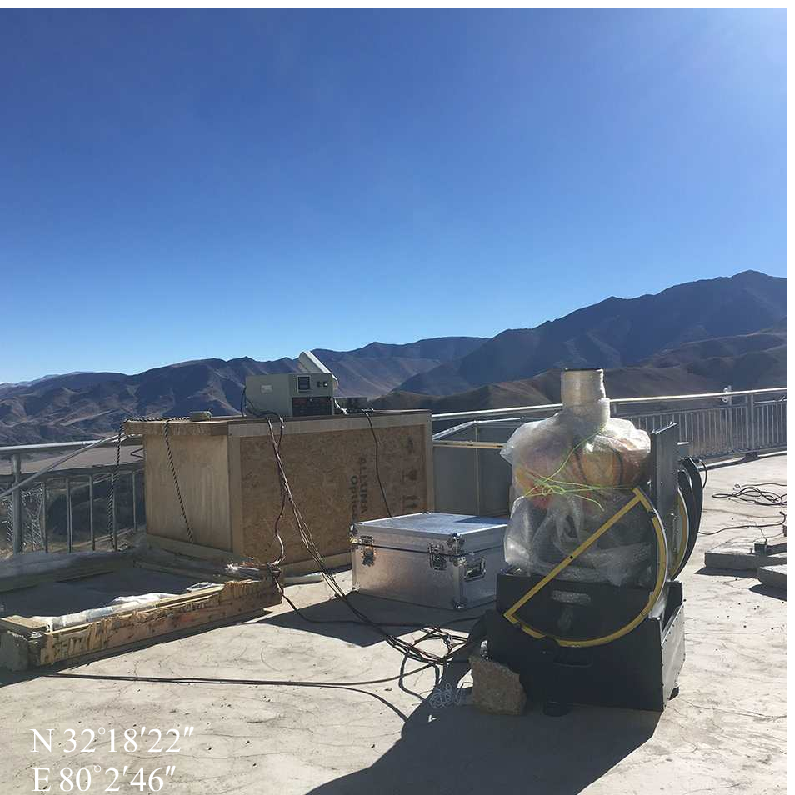}
		\caption{{\small The experimental scene at Ali site.} }\label{Fig:ali_xianchang}
	\end{minipage}%
	\begin{minipage}[t]{0.495\textwidth}
		\centering
		\includegraphics[width=70mm]{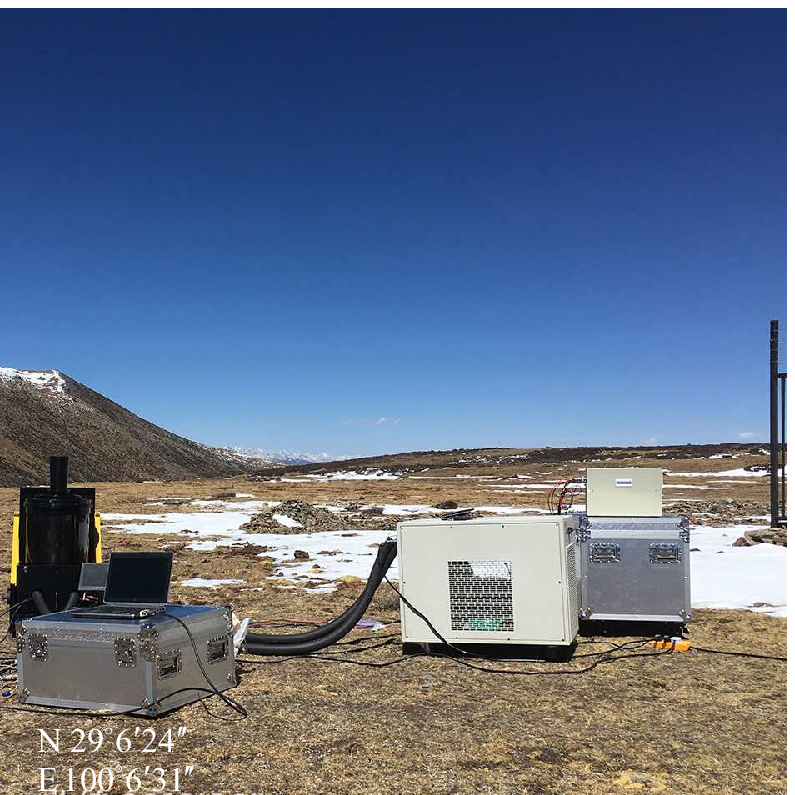}
		\caption{{\small The experimental scene at Daocheng site.}}\label{Fig:daocheng_xianchang}
	\end{minipage}%
\end{figure}

Between 2015 and 2017 an atmospheric mid-infrared radiation meter (AMIRM) was developed  by Yunnan Observatories  and Kunming Institute of Physics. With the AMIRM, we measured the $\rm M'$-band sky brightness and atmospheric extinction at the 4750 m Daocheng site ($29.107^\circ$ north, $100.109^\circ$ east) in late March 2017,  and at the 5100 m Ali site ($32.306^\circ$ north, $80.046^\circ$ east) in late October 2017. The experimental scenes are shown in Figures (\ref{Fig:ali_xianchang} and \ref{Fig:daocheng_xianchang}).
The specifications of the AMIRM are given in Table (\ref{Tab:para_airm}); all optical elements except double sealing windows of the equipment were cooled to $-40^\circ \rm C$. 
\begin{table}[h]
	\begin{center}
		\caption[]{ The specifications of AMIRM.}\label{Tab:para_airm}
		%%Please Capitalize the First Letter of Each Notional Word in table's caption
		\begin{tabular}{clc}
			\toprule
			Equipment & Parameter & Value\\
			\midrule
			\multirowcell{4}{Optical system} & Aperture ($\rm cm$) & 7.5\\
			&Focal length ($\rm cm$)& 15\\
			&Operating temperature ($^\circ \rm C$) & $-40\pm0.02$\\
			&Filter spectral response ($\upmu \rm m$)& $4.605\sim 4.755$\\
			\cmidrule{2-3}
			\multirowcell{3}{HgCdTe detector}& Operating temperature ($^\circ \rm C$)&-196.15\\
			&Spectral response ($\upmu \rm m$)& $3.7\sim 4.8$\\
			&Pixel pitch ($\upmu \rm m$)& 30 \\
			&Format (pixel)& $320\times256$\\
			% new variable
			\bottomrule
		\end{tabular}
	\end{center}
\end{table}

Our filter lies within the spectral range of the $\rm M'$($4.57\upmu \rm m\sim4.79\upmu \rm m$) filter at Mauna Kea Observatories (MKO, \citealt{Leggett2003}).  Hereafter, $\rm M'$ means the spectral response $4.605\upmu \rm m\sim 4.755\upmu \rm m$.

\subsection{Multivariate calibration of the AMIRM}
\label{subsect:cali}

\begin{figure}[htb]
	\begin{minipage}[t]{0.495\linewidth}
	\centering
	\includegraphics[width=65mm]{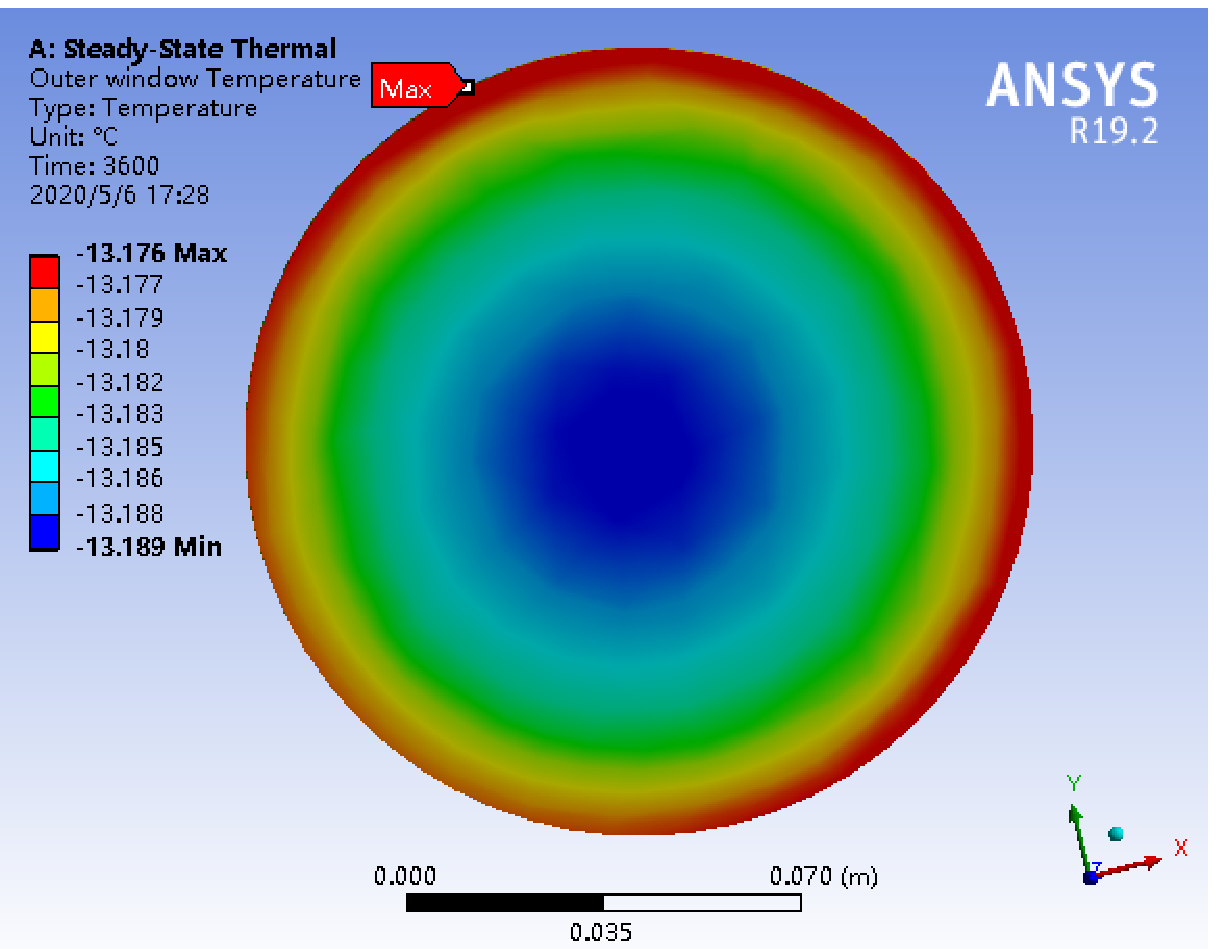}
	\caption{{\small Nephogram of outer window temperature.} }\label{window1_m10c}
    \end{minipage}%	
	\begin{minipage}[t]{0.495\textwidth}
		\centering
		\includegraphics[width=75mm]{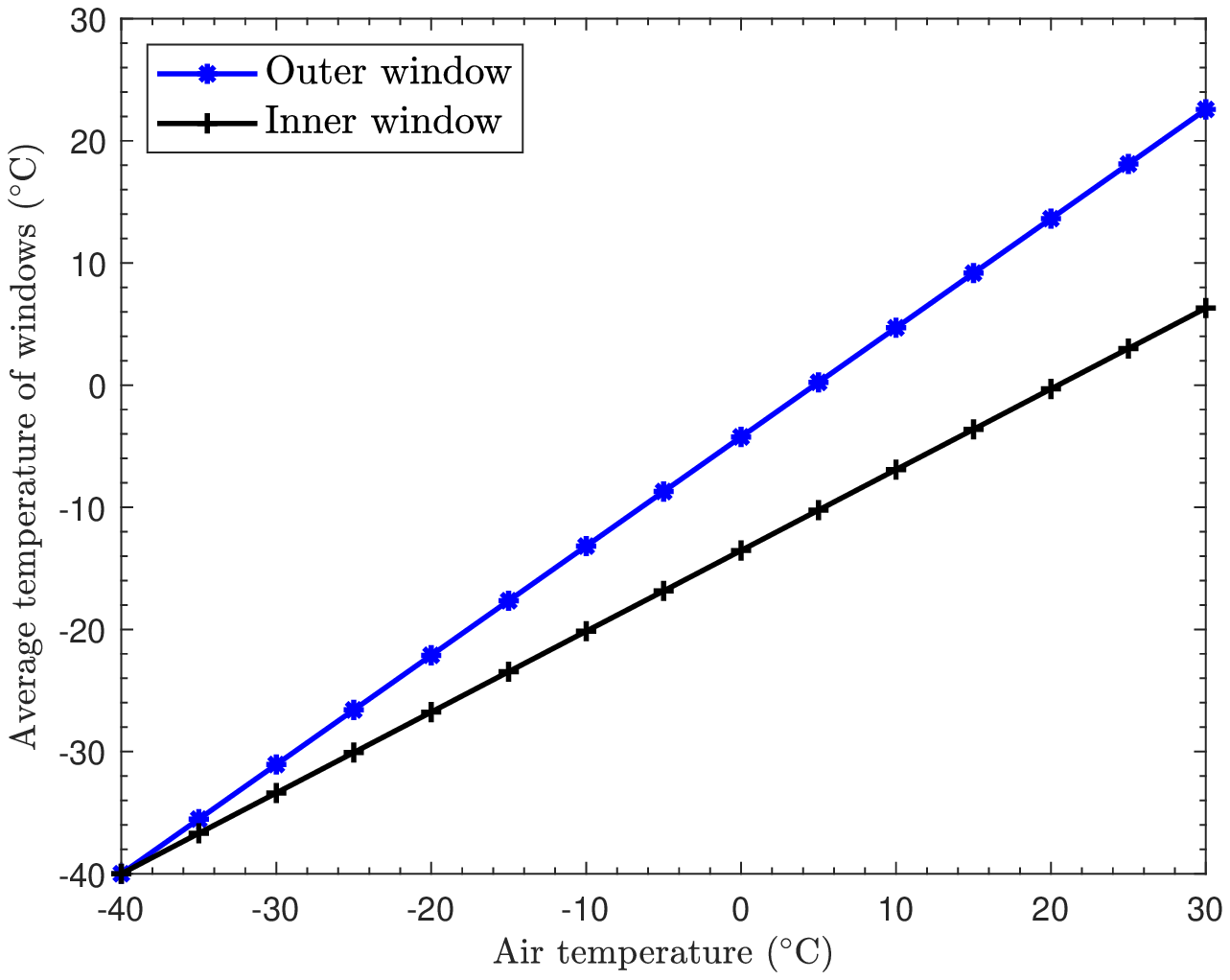}
		\caption{{\small Window temperature with air-temperature.}}\label{window_air}
	\end{minipage}%

\end{figure}

The sealing windows of the AMIRM are close to the air and far from the cooler which is at $-40^\circ \rm C$,  so the temperature of the windows is significantly affected by air-temperature. Figure (\ref{window1_m10c}) shows that the simulating temperatures of the outer window are affected by air-temperature of $-10^\circ \rm C$. 

Figure (\ref{window_air}) illustrates that the simulating average temperatures of the windows vary with air temperature and are far higher than $-40^\circ \rm C$ under the usual air-temperature conditions. Hence, calibration is necessary to eliminate the fluctuating thermal radiation of the windows. In the $\rm M'$-band, the multivariate equation  (\citealt{Zhijun2017-phd}) obtained by calibrating in a temperature environmental chamber is shown in Equation (\ref{eq:tri_cali}):
 
\begin{equation}
I_{adu}=4.875 \times 10^5 \cdot t \cdot L_{sky}^{M'}+8.489\times 10^3\cdot t\cdot L_{amb}^{M'}+2.549 \cdot t+707 \label{eq:tri_cali},
\end{equation}
where $I_{adu}$ is the reading (the unit is in ADU) of the AMIRM, $t$ is exposure time,  $L_{sky}^{M'}$ is radiance of sky, $L_{amb}^{M'}$ is black-body radiance at the air-temperature. The rms error of Equation (\ref{eq:tri_cali}) is $4.25\ \rm {ADU}$.

\subsection{The measured sky brightness at the Ali and Daocheng sites}
\label{subsect:bri_dark}
\begin{figure}[htb]
	\begin{minipage}[t]{0.495\linewidth}
		\centering
		\includegraphics[width=79mm]{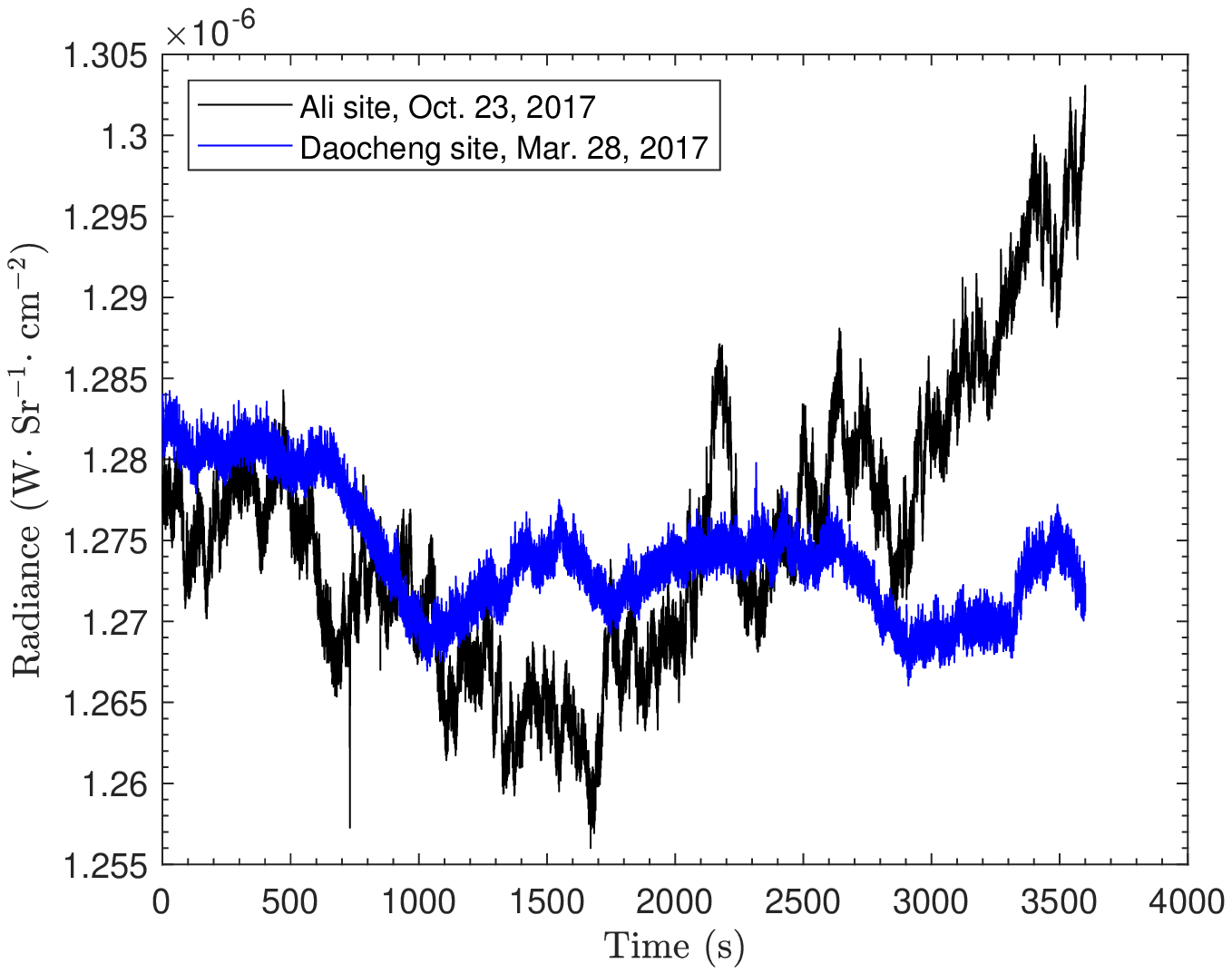}
		\caption{{\small Time sequence of sky brightness at two sites.} }\label{Fig:skynight}
	\end{minipage}%
	\begin{minipage}[t]{0.495\textwidth}
		\centering
		\includegraphics[width=79mm]{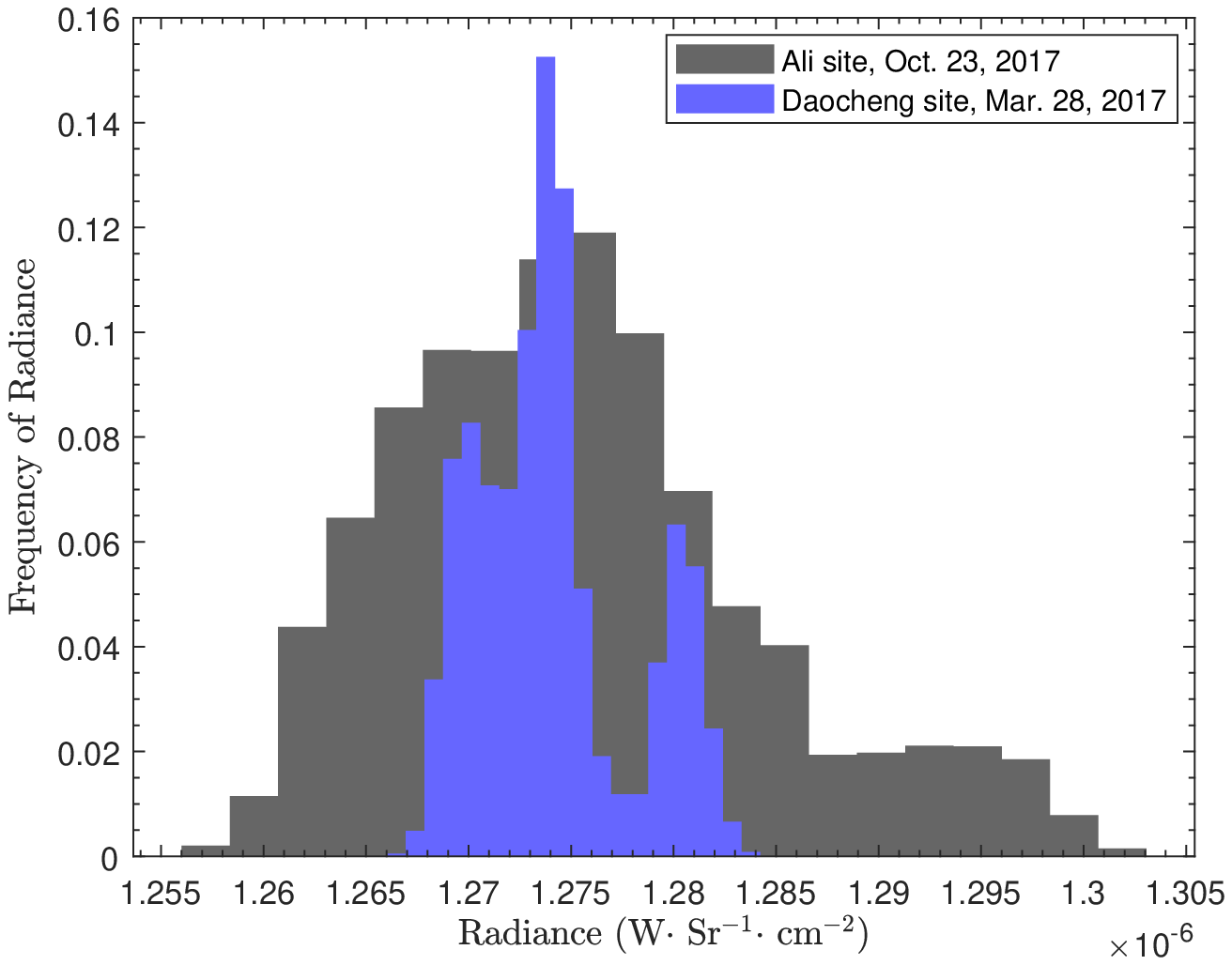}
		\caption{{\small Histogram of sky brightness at two sites.}}\label{Fig:hist_night}
	\end{minipage}%
\end{figure}

In Figure (\ref{Fig:skynight}) we illustrate a one-hour sequence of the $\rm M'$-band at the darkest sky brightness of the zenith which had been measured from 21:35 (6:00) to 22:35 (7:00) Beijing time at Ali (Daocheng). The low frequency fluctuations of sky brightness are the sky noise usually subtracted by chopping technique in the infrared astronomical observations.

The histogram for the above time sequence  is shown in Figure (\ref{Fig:hist_night}), and the mean, standard deviation, minimum, and maximum are shown in Table (\ref{Tab:skybrightness_2site}). The $\rm M'$-band sky brightness at Mauna Kea is also listed as a reference in Table (\ref{Tab:skybrightness_2site}), which is calculated by us based on the simulated data of spectral (between $\rm 4.605\upmu m$ and $\rm 4.755\upmu m$) emission of the sky  on Gemini Observatory's website\footnote{\url{https://www.gemini.edu/observing/telescopes-and-sites/sites\#IRSky}}(\citealt{Lord1992}). 

\begin{table}[htb]
	\begin{center}
		\caption[]{Statistics of the one-hour sequence of $\rm M'$-band sky brightness.}\label{Tab:skybrightness_2site}
		%%Please Capitalize the First Letter of Each Notional Word in table's caption
		\begin{tabular}{lcccc}
			\toprule
			\multirow{2}*{Site}   & \multicolumn{4}{c}{Value ($\times 10^{-6}\ \rm{W\cdot Sr^{-1}}\cdot cm^{-2}$)} \\
			\cmidrule{2-5}
			& Maximum  &  Minimum & Mean     & Standard deviation\\
			\midrule
			Ali                  &1.303&1.256&1.275&0.009 \\  %new variable
			Daocheng             &1.284&1.266&1.274&0.004\\
			Mauna Kea*            &--&--&2.670&--\\
			\bottomrule
		\end{tabular}\\
	   \vspace{1.2ex}\hspace{-0em}
		\footnotesize{*Calculated by using the data with a water vapour column of 1.0 mm and an air mass of 1.0}
	\end{center}
\end{table}

The spectral flux of the $\rm M'$-band zero magnitude is $2.2\times 10^{-15}\ \rm{W\cdot cm^{-2} \cdot \upmu m^{-1}}$ (\citealt{Lena2012}). Thus, the magnitude per square arc-second of the mean sky brightness in Table (\ref{Tab:skybrightness_2site}) is 2.867 $\rm mag\cdot arcsec^{-2}$ at Ali, 2.868 $\rm mag\cdot arcsec^{-2}$ at Daocheng, and 2.064 $\rm mag\cdot arcsec^{-2}$ at Mauna Kea.

\subsection{The measured atmospheric extinction at the Ali and Daocheng sites}
\label{subsec:atmext}

The apertures of equipment used to measure the thermal infrared  atmospheric radiation are usually very small. Therefore, it is generally impossible to measure atmospheric extinction using infrared standard stars in astronomical site surveys. For this reason we present a convenient method for the thermal infrared-band to measure extinction based on the atmospheric radiation transfer equation, as shown in  Equation (\ref{eq:atm_extinc}) (\citealt{Zhijun2018,Wang2020}):

\begin{equation}
I_{sky}=a\cdot(1-e^{-\overline{O}_d\cdot AM}),
\label{eq:atm_extinc}
\end{equation}
where $I_{sky}$ is ADU readings of $\rm M'$-band sky calibrated by the Equation (\ref{eq:tri_cali}), $AM$ is air mass at any zenith angles, and $\overline{O}_d$ is the average optical depth in $\rm M'$-band at the zenith. 

\begin{figure}[htb]
	\begin{minipage}[t]{0.495\linewidth}
		\centering
		\includegraphics[width=80mm]{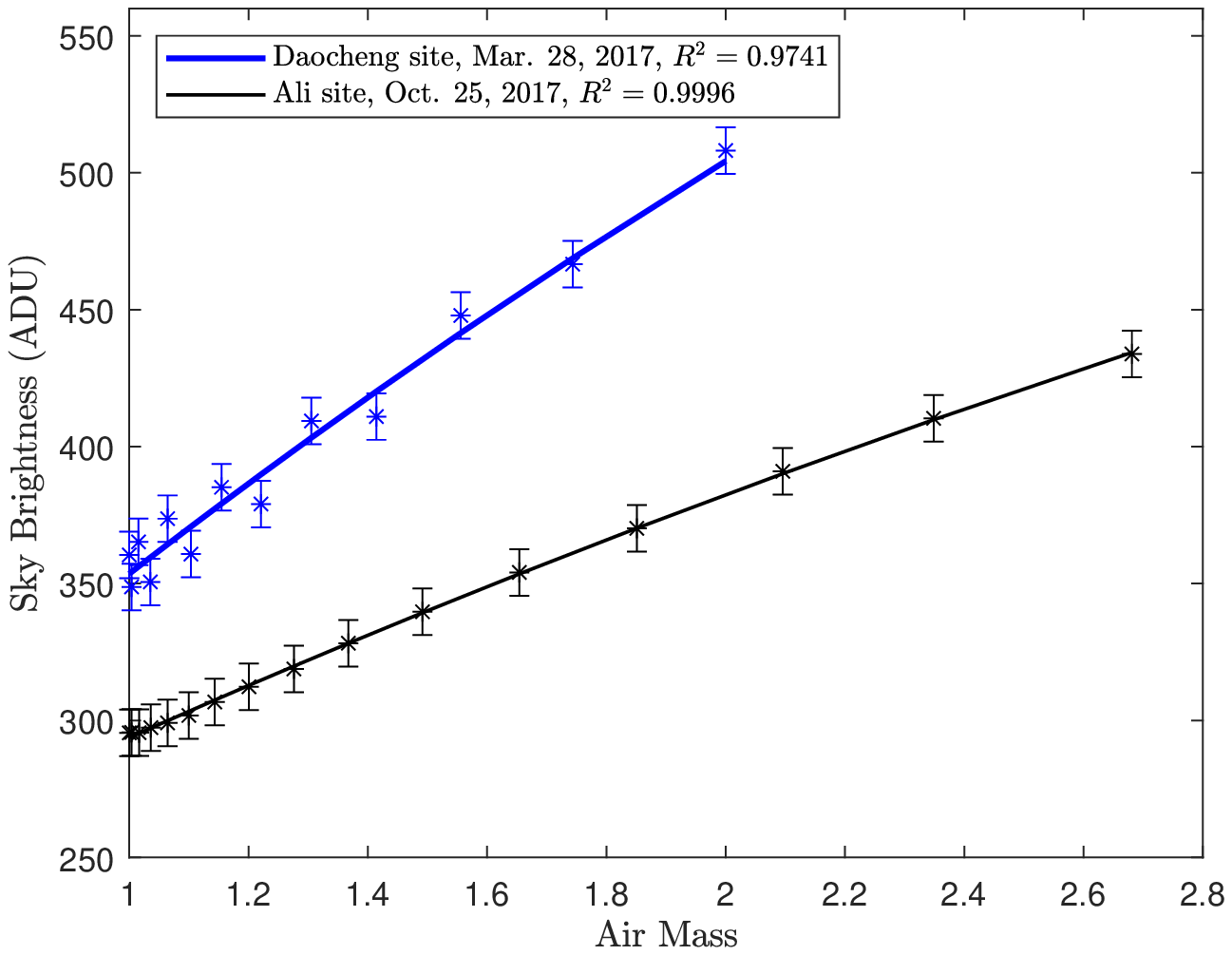}
		\caption{{\small Fitting sky brightness at different air mass.}}\label{fig:tr_measure}
	\end{minipage}%
	\begin{minipage}[t]{0.495\textwidth}
		\centering
		\includegraphics[width=80mm]{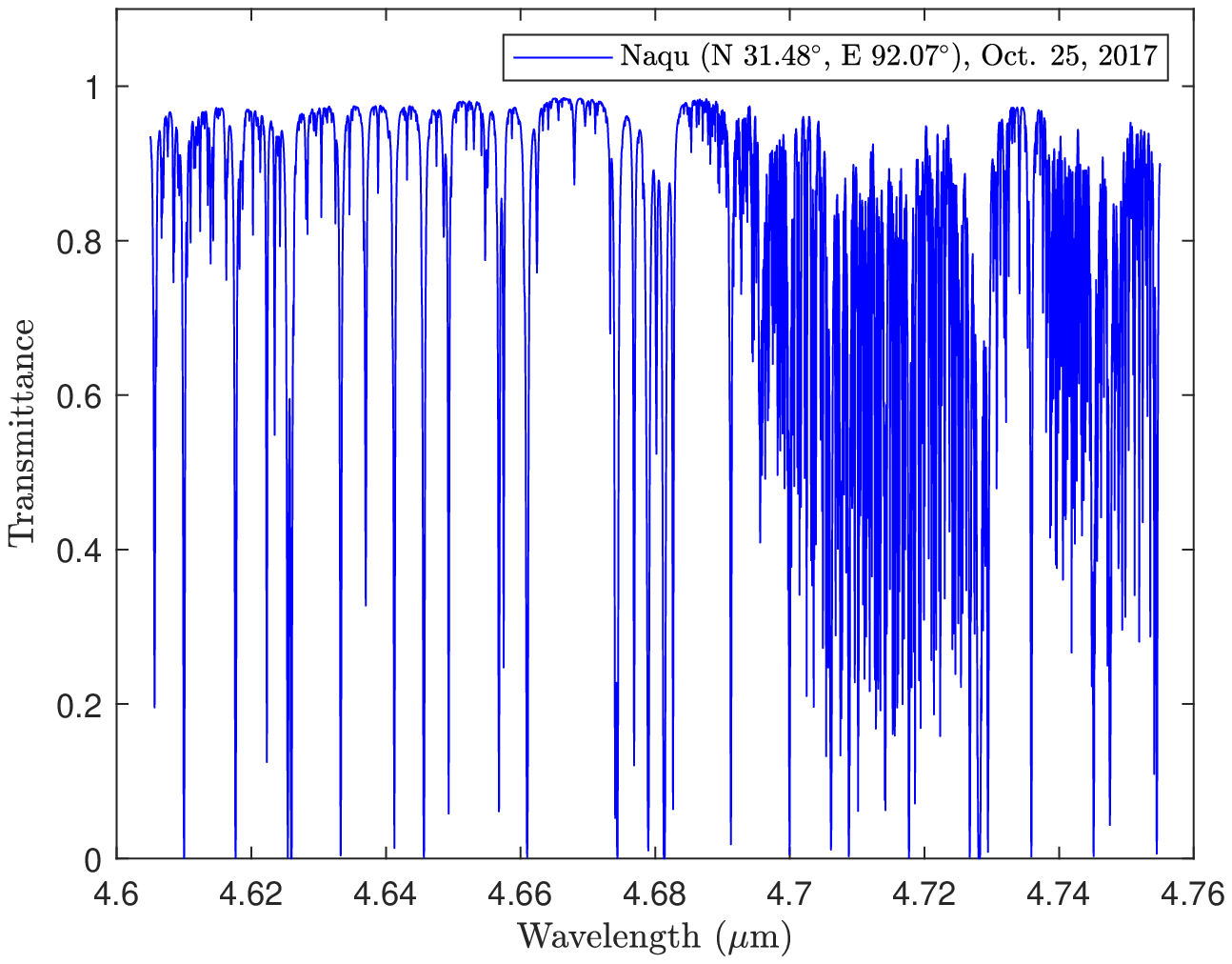}
		\caption{{\small Transmittance calculated by LBLRTM.} }\label{fig:tr_lblrtm}
	\end{minipage}%
\end{figure}

Figure (\ref{fig:tr_measure}) shows that the measured sky brightness at different air masses is fit by Equation (\ref{eq:atm_extinc}), and the $\overline{O}_d$ obtained by fitting is 0.18 at Ali, and 0.23 at Daocheng.  From $e^{-\overline{O}_d\cdot {AM}}$, the transmittance at the zenith is 0.84 at Ali and 0.80 at Daocheng.  As a reference, by using LBLRTM (Line-By-Line Radiative Transfer Model), we calculate the spectral transmittance at the zenith with radiosondes data of Naqu\footnote{\url{http://weather.uwyo.edu/upperair/sounding.html}. \color{blue}{The Radiosonde data of Ali and Daocheng are lacking.}}, and the transmittance results  are shown in Figure (\ref{fig:tr_lblrtm}). By integrating the spectral transmittance in $\rm M'$-band, the mean transmittance is 0.82 at Naqu.

The atmospheric extinction can be calculated by 

\begin{equation}
\Delta m_{ext}=m_{atm}-m_{exoatm}=-2.5{\rm log}_{10}(e^{-\overline{O}_d \cdot{AM}})\approx1.086\times \overline{O}_d \cdot{AM},
\label{eq:atm_ext}
\end{equation}
where $m_{atm}$ is the magnitude through the atmosphere, and $m_{exoatm}$ is the magnitude beyond the atmosphere.
Atmospheric extinctions at a unit air mass calculated by Equation (\ref{eq:atm_ext}) are given in Table (\ref{Tab:atm_ext_3site}).

\begin{table}[htb]
	\begin{center}
		\caption[]{The $\rm M'$-band atmospheric extinction.}\label{Tab:atm_ext_3site}
		\begin{tabular}{lcc}
			\toprule
			Site      &    \makecell{Atmospheric extinction at unit air mass \\($\rm {mag} \cdot \rm{airmass}^{-1}$)}\\
			\midrule
			Ali       &  $0.20$\\
			Daocheng  &  $0.25$\\
			Mauna Kea (\citealt{Leggett2003}) &  $0.23$\\
			\bottomrule
		\end{tabular}
	\end{center}
\end{table}

The results of atmospheric extinction at Ali and Daocheng are slightly different from MKO because the wave-band of our filter is narrower than MKO's.

%%%%%%%%%%%%%%%
\section{The sensitivity results and discussion}
\label{sect:lm}
In this section we first test the sensitivity model by calculating the $\rm M'$-band limiting magnitudes of 10 m telescopes at the above three sites and then discuss the problem of detector selection and instrumental thermal-control. The involved input parameters are listed in Table (\ref{Tab:para_s2n}), and two infrared detectors with different levels are chosen for comparison.  $Dark=0.1e^-\cdot \rm s^{-1}\cdot\rm{pixel}^{-1}$ and $Ron=10e^-$ are parameters of the scientific detector used on Keck $\rm{\luoma{2}}$ (\citealt{Mclean2003}). $Dark=10^{5}e^-\cdot \rm s^{-1}\cdot\rm{pixel}^{-1}$ and $Ron=500e^-$ represent the noise level of commercial infrared detectors\footnote{\url{https://www.lynred.com/products}} (\citealt{Rubaldo2016Recent}). The required site parameters are from Tables (\ref{Tab:skybrightness_2site} and \ref{Tab:atm_ext_3site}).

\begin{table}[h]
	\begin{center}
		\caption[]{ Input parameters for calculating $\rm M'$-band $S/N$ and magnitudes.}\label{Tab:para_s2n}
		%%Please Capitalize the First Letter of Each Notional Word in table's caption
		\begin{tabular}{clcc}
			\toprule
			Type of parameter & Parameter name  & \multicolumn{2}{c}{Value}\\
			\midrule
			\multirowcell{2}{Observational \\parameters} & Total exposure time ($\rm s$) & \multicolumn{2}{c}{1}\\
			
			&Elementary exposure time (s)                       &   \multicolumn{2}{c}{$1.6\times10^{-5}\sim2.6\times10^{-1}$ @ $70\%$ well fill}\\
			\cmidrule{2-4}
			\multirowcell{5}{Equipment\\ specifications}&Telescope+Relay optics transmittance & \multicolumn{2}{c}{Photometry: 0.5}\\ 
			&Quantum efficiency of detector & \multicolumn{2}{c}{0.85}\\ 
			&Full-well capacity of detector ($e^{-}$)      &  \multicolumn{2}{c}{$\sim10^6$}\\ 
			&Dark electrons ($e^-\cdot \rm s^{-1}\cdot\rm{pixel}^{-1}$) &  $10^{-1}$ @ $-243^\circ \rm C$& $10^{5}$ @ $-143^\circ \rm C$\\
			&RMS readout noise ($e^-\cdot\rm{pixel}^{-1}$)           & 10 &  500 \\
			 % new variable
			\bottomrule
		\end{tabular}
	\end{center}
\end{table}

\subsection{$\rm M'$-band limiting magnitudes of 10 m telescopes}
\label{subsect:lm3sigma}

\begin{figure}[htb]
	\begin{minipage}[t]{0.495\textwidth}
		\centering
		\includegraphics[width=75mm]{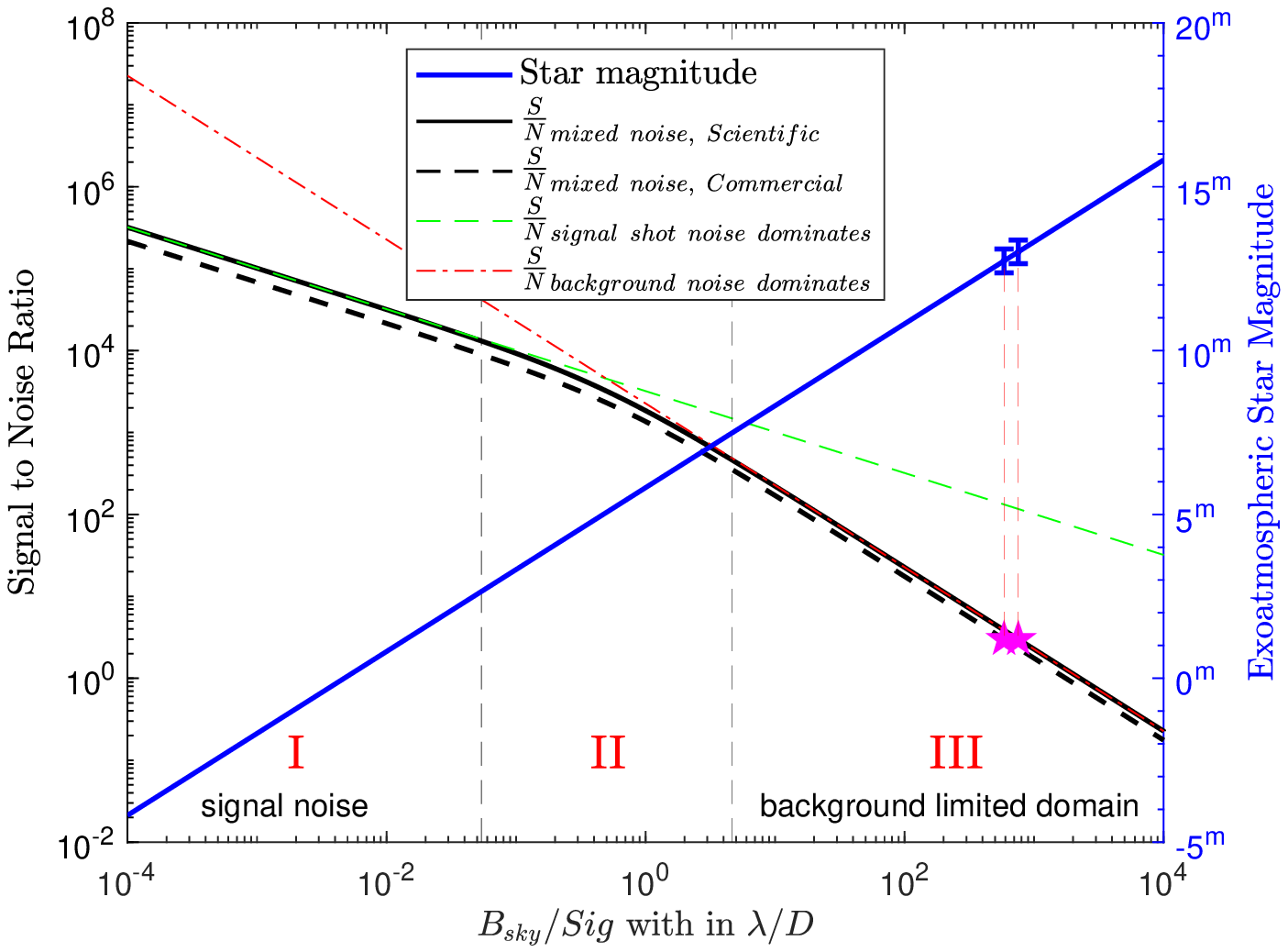}
		\caption{{\small $S/N$ and magnitudes of 10m telescopes at Ali.} }\label{Fig:s2ncurve}
	\end{minipage}%
	\begin{minipage}[t]{0.495\linewidth}
		\centering
		\includegraphics[width=75mm]{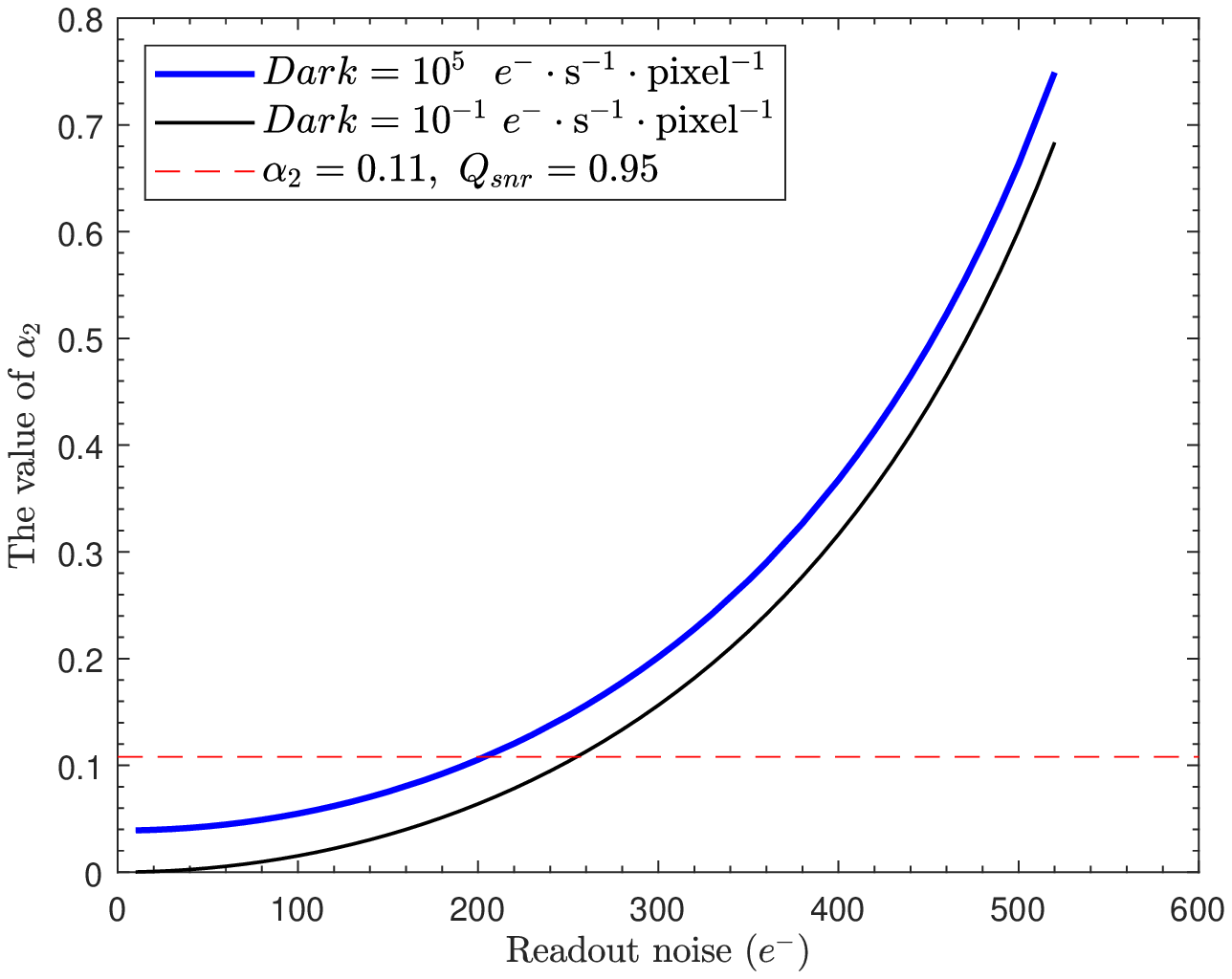}
		\caption{{\small The effects of detector noise at Ali.}}\label{fig:alpha2}
	\end{minipage}%
\end{figure}

Under the circumstances that an infrared telescope is used at its diffraction limit, we have that $\Omega_{star}=\pi(1.22\lambda/D)^2$. According to Equation (\ref{eq:bgd_sig1}), the electrons per second at a unit pixel from the sky background at Ali, Daocheng, and Mauna Kea are about $2.567\times10^6$, $2.565\times10^6$, and $5.376\times10^6$ respectively. If an airy disk is resolved at the Nyquist frequency, $n_{pixel}\approx 4$ and $\alpha_2=4\times(Dark\cdot t+Ron^2)/(\tau_{ins}\eta L_{sky}\pi^2(0.61\lambda)t\frac{\lambda}{hc})$. The limiting magnitude at $S/N=3$ is denoted by $m_{3\sigma}$. Under the condition that the instrumental emission is ignored ($\alpha_1=0$), we calculate the $m_{3\sigma}$  by Equations (\ref{eq:s2n} and \ref{eq:mag_3sigma}), which is shown in Table (\ref{Tab:alpha}). The curves that relate $S/N$ and magnitudes at Ali are shown in Figure (\ref{Fig:s2ncurve}), where the two five-pointed stars mean $S/N=3$.

\begin{table}[htb]
	\begin{center}
		\caption[]{Limiting magnitudes of 10m telescopes without instrumental emission.}\label{Tab:alpha}
		\begin{tabular}{lllllllr}
			\toprule
			\multirowcell{2}{Site} & \multicolumn{3}{c}{Scientific detector}     &\multicolumn{3}{c}{Commercial detector} & \multirowcell{2}{$m_{3\sigma}^{\rm BLCAS}$}\\
			\cmidrule(lr){2-4}\cmidrule(lr){5-7}
			                & $\alpha_2$&  $Q_{snr}$ & $m_{3\sigma}$    & $\alpha_2$&  $Q_{snr}$ & $m_{3\sigma}$&    \\
			\midrule
			 Ali     &$1.5026\times10^{-4}$& 0.9999  & $13.0122^{\rm m}$ &0.6633 &0.7755 &$12.7361^{\rm m}$&$13.0123^{\rm m}$\\
			 Daocheng      & $1.5026\times10^{-4}$ &0.9999 &$12.9596^{\rm m}$ & 0.6634  &0.7755   &$12.6835^{\rm m}$&$12.9597^{\rm m}$\\
			 Mauna Kea  &$1.5018\times10^{-4}$ & 0.9999 & $12.5716^{\rm m}$ &0.6305 &0.7832 &$12.3063^{\rm m}$&$12.5716^{\rm m}$\\			
		\bottomrule
		\end{tabular}\\
	\end{center}
\end{table}

In Table (\ref{Tab:alpha}),  $m_{3\sigma}^{\rm BLCAS}$ is $3\sigma$ magnitude limited by the atmosphere. The scientific detector has little impact on the limiting magnitude, which is less than $0.0001^{\rm m}$ at all three sites. The commercial detector has a significant impact on the limiting magnitude, which is about $0.3^{\rm m}$. 

As discussed in Section (\ref{subsec:qsnr}), for $Q_{snr}\ge0.95$, the calculated $3\sigma$ magnitudes are greater than $12.956^{\rm m}$, $12.904^{\rm m}$, and $12.516^{\rm m}$ at Ali, Daocheng, and Mauna Kea respectively, which decrease at most by $0.056^{\rm m}$ relative to $m_{3\sigma}^{\rm BLCAS}$. Obviously, the $Q_{snr}$ can clearly reflect whether the sky background limited sensitivity is approached. Hence, based on the requirements of observations, we can easily select appropriate detectors using $Q_{snr}$. Figure (\ref{fig:alpha2}) shows the $\alpha_2$ varying with different detector noise and the curve of $\alpha_2=0.11$ determined by $Q_{snr}=0.95$.

From Figure (\ref{fig:alpha2}) we should select the detector such that the readout noise is less than $250e^-$ for $Dark=10^{-1} e^-\cdot \rm s^{-1}\cdot pixel^{-1}$, and $200e^-$ for $Dark=10^{5} e^-\cdot \rm s^{-1}\cdot pixel^{-1}$. The dark current just provides a small contribution to $\alpha_2$ because it is far less than the sky background electrons (about $2.6\times10^6 e^-\cdot \rm s^{-1}\cdot pixel^{-1}$) in $\rm M'$-band for the above two detectors. The dark current  significantly decreases  with decreasing the temperature of detectors. Therefore, for deep cryogenic detectors, we should pay special attention to readout noise.

\subsection{The discussion on noise distribution and instrumental thermal-control}
\label{subsect:noise_thermal}

For thermal infrared observations, in addition to detector noise, we should also strictly control instrumental thermal emission, which is determined by the real emissivity  and temperature of the instruments. By using the above analysis, we can select a deep cryogenic detector such that the dark current can be ignored  and the readout noise is $200e^-$.  According to the measured sky brightness in Table (\ref{Tab:skybrightness_2site}) and Equation (\ref{eq:eff_real}), the effective emissivity of sky is 0.1554 of $-5.4^\circ \rm C$ at Ali, and 0.1898 of $-10^\circ \rm C$ at Daocheng. In order to meet $Q_{snr}\ge0.95$, based on Equation (\ref{eq:alpha1}) and the readout noise of $200e^-$, the effective emissivity of instruments should be less than 0.003 at Ali, and less than 0.004 at Daocheng. Obviously, if the real emissivity is equal to the effective emissivity, it is not necessary to refrigerate the instruments. For three instruments with given real emissivity, the required cooling temperatures are also calculated by Equation (\ref{eq:eff_real}), which are shown in Table (\ref{Tab:cool_tem}).
\begin{table}[htb]
	\begin{center}
		\caption[]{The required cooling temperature of instruments for $Q_{snr}=0.95$.}\label{Tab:cool_tem}
		\begin{tabular}{rrrr}
			\toprule
			\multirowcell{2}{Site} & \multicolumn{3}{c}{The cooling temperature ($^\circ \rm C$)}\\
			\cmidrule(lr){2-4}
			&$\epsilon_{ins}=0.01$&$\epsilon_{ins}=0.25$ & $\epsilon_{ins}=0.5$\\
			\midrule
			Ali @ $-5.4^\circ \rm C$ &$-29.47$&$-79.03$&$-87.18$\\
			Daocheng @ $-10^\circ \rm C$  &$-29.50$&$-79.05$&$-87.19$\\
			\bottomrule
		\end{tabular}\\

	\end{center}
\end{table}

In Table (\ref{Tab:cool_tem}), all elements have the same cooling temperatures for ease of calculation. However, the open-air elements of a telescope are hard to cool down to the same temperature as the relay optics. Therefore, the relay instrument should be cooled down to a lower temperature than in Table (\ref{Tab:cool_tem}), and the real emissivity of the telescope should be kept as low as possible.

\begin{figure}[htb]
	\begin{minipage}[t]{0.495\linewidth}
		\centering
		\includegraphics[width=80mm]{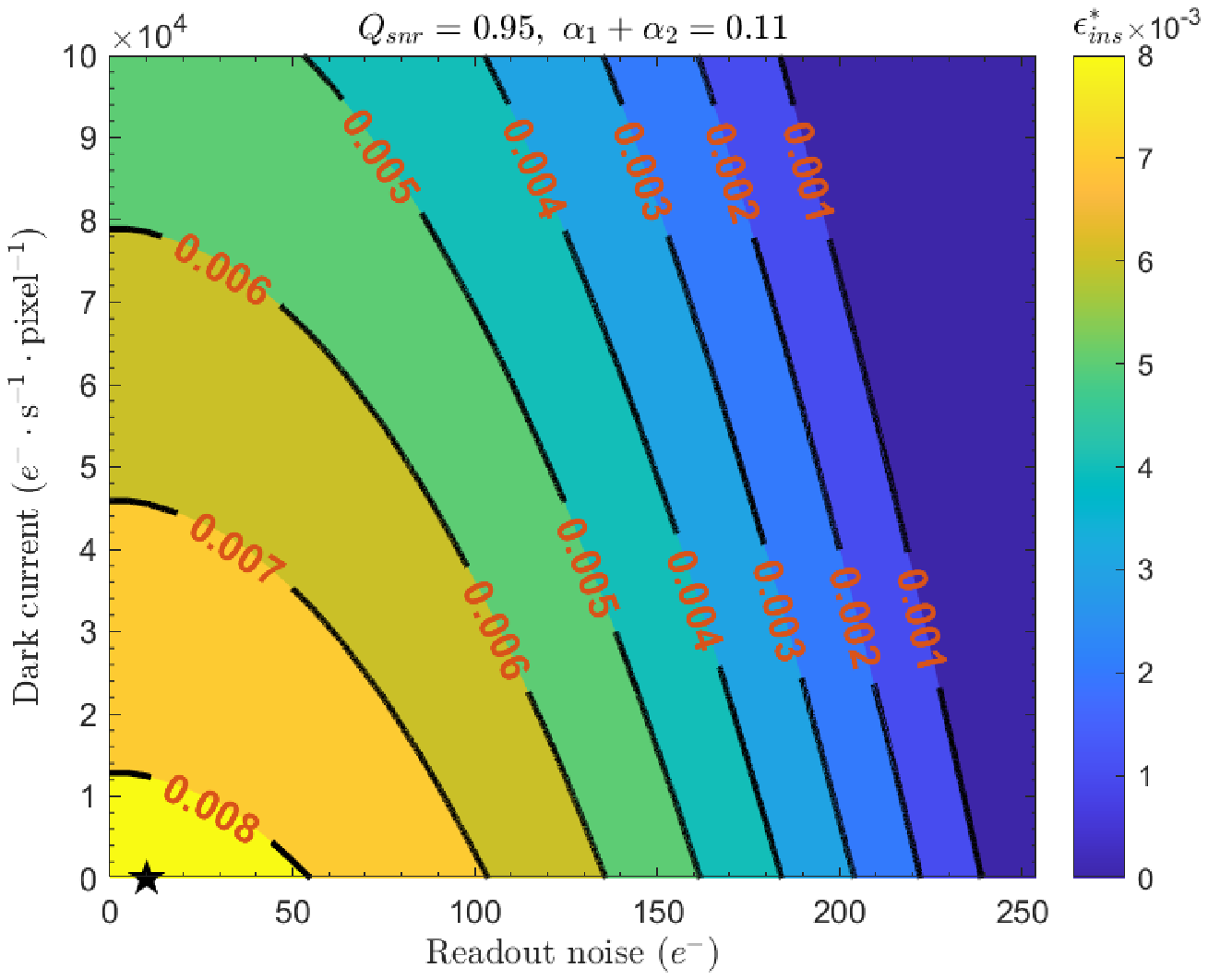}
		\caption{{\small The distribution of noise between  a detector and instruments at Ali.}}\label{fig:alpha21}
	\end{minipage}%
	\begin{minipage}[t]{0.495\textwidth}
		\centering
		\includegraphics[width=80mm]{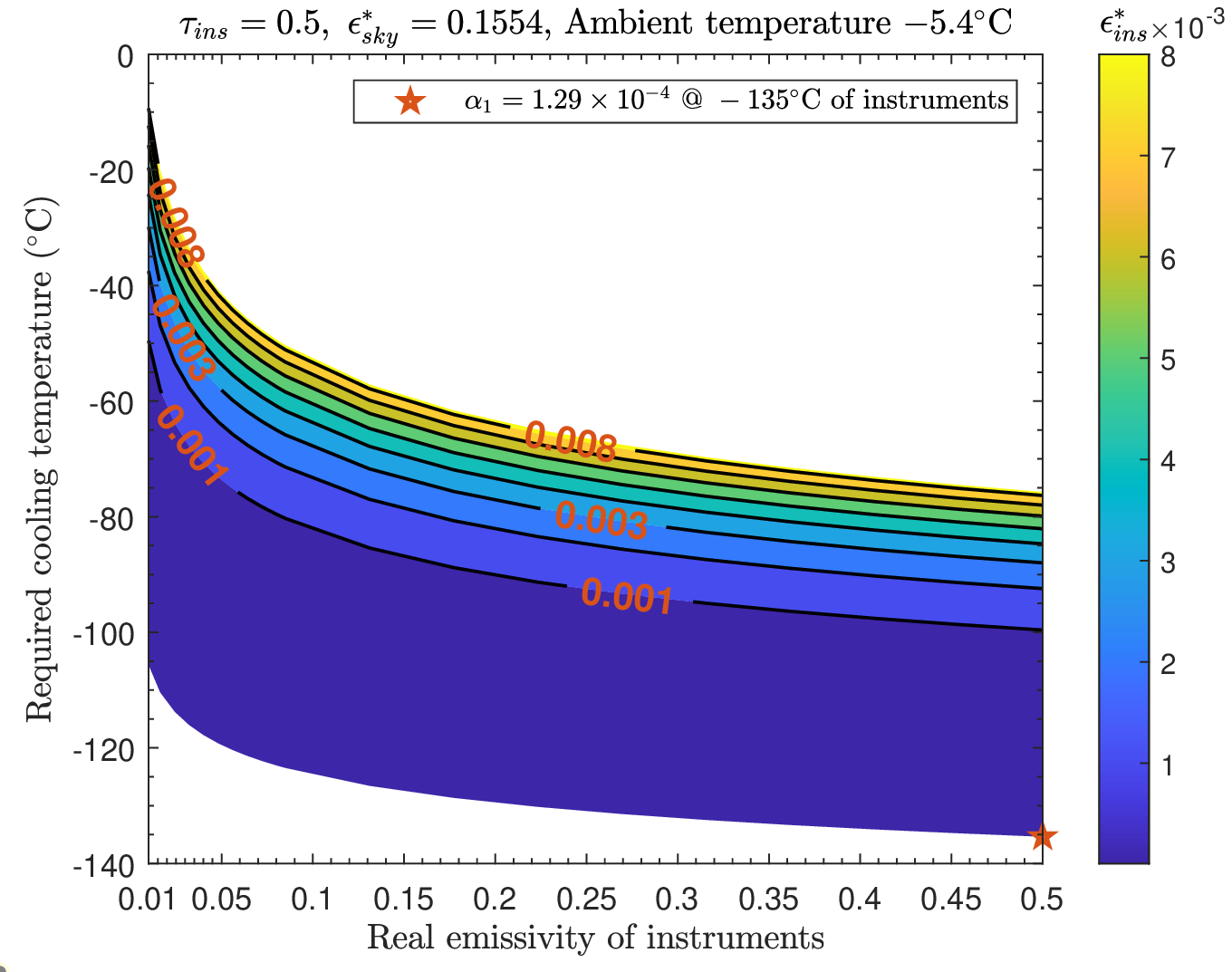}
		\caption{{\small The thermal-control of instruments at Ali.} }\label{Fig:thermal_con}
	\end{minipage}%
\end{figure}

By taking the case of Ali site we calculate more possible combinations  of detector noise and instrument noise, which are shown in Figure (\ref{fig:alpha21}). The required cooling temperatures of instruments that vary with the instrumental real emissivity are calculated by Equation (\ref{eq:eff_real}), which are shown in Figure (\ref{Fig:thermal_con}).
 
The black five-pointed star in Figure (\ref{fig:alpha21}) presents the noise of the scientific detector on Keck $\rm{\luoma{2}}$. The red-brown  five-pointed star in Figure (\ref{Fig:thermal_con}) demonstrates that the instrumental thermal noise can be ignored for the instruments cooled down to $-135^\circ \rm C$. From the above two figures, when the thermal emission of instruments is strictly controlled, we are allowed to choose a less-than-ideal detector whose maximum readout noise is about $250e^-$.  In other words an excellent thermal-control design of instruments makes it easier to select a required detector.

\section{Conclusions}
\label{sect:conclusion}

In this paper we constructed a sensitivity model used for the guidance of large telescope site surveys, detector selection, and instrumental thermal-control. The model was tested with the measured $\rm M'$-band sky brightness and atmospheric extinction in western China, and then the sensitivities limited by the atmosphere for 10 m telescopes were estimated.  According to a given requirement of observational sensitivity, we discussed the principles of noise distribution and the problem on instrumental thermal-control. The better a site is, the more one needs to choose an excellent detector and strictly control instrumental thermal emission. When the instrument noise can be ignored, readout noise should still be less than $250e^-$.  For  Ali or Daocheng, if we choose a detector such that the readout noise is $200e^-$ and the dark current can be ignored, in order to fulfill  $Q_{snr}\ge0.95$, the instruments whose real emissivity is 0.5 should be cooled to below $-87.2^\circ \rm C$.

If the statistical results of long-term data on sky brightness and atmospheric extinction are obtained, the model may be used to evaluate more objectively the sensitivity of large telescopes in a given infrared-band  at any site and to guide the detector selection and the instrumental thermal-control effectively.

\begin{acknowledgements}
This work was funded by the National Natural Science Foundation of China (NSFC)
under No.11803089, No.U1931124.
\end{acknowledgements}

%\appendix                  %%appendicial material is supported
%
%\section{This shows the use of appendix}
%A postscript file is actually an ASCII text file (you may even edit it).
%However, you need to transfer a PDF file or any compressed or packaged
%file in binary mode when using FTP.
%
%\section{What is SCI?}
%SCI is the abbreviation of Science Citation Index system powered by
%the Institute for Scientific Information (ISI). For details please
%visit {\it http://apps.isiknowledge.com}.

\bibliographystyle{raa}
\bibliography{bibtexzhaozhijun}

\begin{thebibliography}{25}
\providecommand\natexlab[1]{#1}
\providecommand\JournalTitle[1]{#1}

\bibitem[Ashley {et~al.}(1996)]{Ashley1996}
Ashley, M.~C., Burton, M.~G., Storey, J.~W., {et~al.} 1996, PASP, 108,
  721–723

\bibitem[Cui {et~al.}(2018)]{Cui2018}
Cui, X., Zhu, Y., Liang, M., {et~al.} 2018, in Ground-based and Airborne
  Telescopes VII

\bibitem[Deng {et~al.}(2016)]{Deng2016}
Deng, Y., Liu, Z., Qu, Z., Liu, Y., \& Ji, H. 2016, in Astronomical Society of
  the Pacific Conference Series, Vol. 504, 203

\bibitem[Dong {et~al.}(2018)]{Dong2018Design}
Dong, S.-c., Wang, J., Tang, Q.-j., {et~al.} 2018, Review of Scientific
  Instruments, 89, 023107

\bibitem[Feng {et~al.}(2020)]{Feng2020Site}
Feng, L., Hao, J.-X., Cao, Z.-H., {et~al.} 2020, RAA, 20, 80

\bibitem[Leggett {et~al.}(2003)]{Leggett2003}
Leggett, S.~K., Hawarden, T.~G., Currie, M.~J., {et~al.} 2003, MNRAS, 345, 144

\bibitem[Lena {et~al.}(2012)]{Lena2012}
Lena, P., Rouan, D., Lebrun, F., {et~al.} 2012, Observational astrophysics
  (Springer Berlin Heidelberg)

\bibitem[Liu {et~al.}(2018)]{Liu2018}
Liu, Y., Li, X., Song, T., Zhang, X., \& Song, Q. 2018, in Society of
  Photo-Optical Instrumentation Engineers (SPIE) Conference Series, Vol. 10704,
  1070422

\bibitem[Liu {et~al.}(2012)]{Liu2012}
Liu, Z., Deng, Y., \& Ji, H. 2012, Proceedings of Spie the International
  Society for Optical Engineering, 8444, 213

\bibitem[Lord(1992)]{Lord1992}
Lord, S.~D. 1992, NASA Technical Memorandum, 103957

\bibitem[Mclean(2003)]{Mclean2003}
Mclean, I.~S. 2003, Proceedings of Spie the International Society for Optical
  Engineering, 4834, 111

\bibitem[Phillips {et~al.}(1999)]{Phillips1999}
Phillips, A., Burton, M.~G., Ashley, M. C.~B., {et~al.} 1999, ApJ, 527, 1009

\bibitem[Qian {et~al.}(2015)]{Qian2015Numberical}
Qian, X., Yao, Y., Wang, H., {et~al.} 2015, Journal of Physics Conference, 595,
  012028

\bibitem[Rubaldo {et~al.}(2016)]{Rubaldo2016Recent}
Rubaldo, L., Brunner, A., Guinedor, P., {et~al.} 2016, in Quantum Sensing and
  Nano Electronics and Photonics XIII

\bibitem[Smith \& Harper(1998)]{Smith1998}
Smith, C., \& Harper, D. 1998, PASP, 110, 747

\bibitem[Song {et~al.}(2020)]{Song2020}
Song, T.~F., Liu, Y., Wang, J.~X., Zhang, X.~F., \& Ruan, Y. 2020, RAA, 20, 085

\bibitem[Tang {et~al.}(2018)]{Tang2018}
Tang, Q.-J., Wang, J., Dong, S.-C., Chen, J.-T., \& Tang, P. 2018, Journal of
  Astronomical Telescopes, Instruments, and Systems, 4

\bibitem[Wang {et~al.}(2020)]{Wang2020}
Wang, F.-X., Xu, F.-Y., Guo, J., {et~al.} 2020, RAA, 20, 134

\bibitem[Wang {et~al.}(2013)]{Wang2013}
Wang, H., Yao, Y., \& Liu, L. 2013, Acta Optica Sinica, 33, 0301006

\bibitem[Wang {et~al.}(2018)]{Wang2018}
Wang, J., Zhang, Y.~H., Tang, Q.~J., Dong, S.~C., \& hao Jia, M. 2018, in
  Ground-based and Airborne Telescopes VII

\bibitem[Westphal(1974)]{Westphal1974}
Westphal, J. 1974, NASA Technical Reports, NGR-05-002-185

\bibitem[Wu {et~al.}(2016)]{Wu2016}
Wu, N., Liu, Y., \& Zhao, H. 2016, Acta Astronomica Sinica

\bibitem[Yao(2005)]{Yao2005}
Yao, Y. 2005, Journal of the Korean Astronomical Society, 38, 113

\bibitem[Zhao(2017)]{Zhijun2017-phd}
Zhao, Z.-J. 2017, Research on background radiation characteristics of
  ground-based infrared solar observation, Doctor of philosophy, University of
  Chinese Academy of Sciences

\bibitem[Zhao {et~al.}(2018)]{Zhijun2018}
Zhao, Z.-J., Xu, F.-Y., Wei, C.-Q., \& Yang, K. 2018, Infrared Technology, 40,
  718

\end{thebibliography}

\label{lastpage}

\end{document}